\documentclass{llncs}
    \usepackage{amsmath,amssymb,stmaryrd,amssymb}
    \usepackage{url,xspace,enumerate,epic,eepic,verbatim,hyperref,color}
    \usepackage[dvipsnames]{xcolor}
    \usepackage[utf8]{inputenc}
    \usepackage[final,stretch=10,shrink=20]{microtype}
    \usepackage[scaled=0.85]{helvet}
    \usepackage[scaled=0.82]{beramono}

    \def\mpar#1{\paragraph{\bf#1.}}

    \def\enc#1#2{\textsf{Enc}_{#1}(#2)}
    \def\dec#1#2{\textsf{Dec}_{#1}(#2)}

    \def\pk{\mathit{pk}}
    \def\eenc#1{\textsf{Enc}_{\mathit{pk}_e}(#1)}
    \def\penc#1{\textsf{Enc}_{\mathit{pk}_p}(#1)}
    \def\cenc#1{\textsf{Enc}_{\mathit{pk}_c}(#1)}
    \def\aenc#1{\textsf{Enc}_{\mathit{pk}_a}(#1)}
    \def\prob{\mathsf{Prob}}
    \def\equivnegl{\equiv_{\textsf{negl}} }
    \def\codes{\textsf{Codes}}
    \def\econf{e_\textsf{\itshape{conf}} }
    \def\cfin{c_\textsf{\itshape{fin}} }
    \def\comp{\,\|\,}

    \title{Cast-as-Intended Mechanism with \\  Return Codes Based
    on PETs\\ Extended Version}
    \author{Achim Brelle \and Tomasz Truderung}
    \institute{Polyas GmbH
        \\\email{a.brelle@polyas.de, t.truderung@polyas.de}}

\begin{document}
\maketitle


\begin{abstract}
    We propose a method providing cast-as-intended verifiability
    for remote electronic voting. The method is based on
    plaintext equivalence tests (PETs), used to match the cast
    ballots against the pre-generated encrypted code tables.
    
    Our solution provides an attractive balance of security and
    functional properties. It is based on well-known
    cryptographic building blocks and relies on standard
    cryptographic assumptions, which allows for relatively simple
    security proofs. Our scheme is designed with a built-in
    fine-grained distributed trust mechanism based on
    threshold decryption.  It, finally, imposes only very little
    additional computational burden on the voting platform, which
    is especially important when voters use devices of restricted
    computational power such as mobile phones.  At the same time,
    the computational cost on the server side is very reasonable
    and scales well with the increasing ballot size.
\end{abstract}

\section{Introduction}

    Modern electronic voting systems are expected to provide a
    combination of security guarantees which includes, most
    importantly, ballot secrecy and end-to-end verifiability. For
    the latter, one crucial part is so-called
    \emph{cast-as-intended} verifiability which means that a
    voter has means to make sure that the ballot cast on his or
    her behalf by the voting client application and recorded by
    the voting server contains the intended voting option, as
    chosen by the voter. This property must be guaranteed without
    assuming that the voter platform is honest. Indeed, such
    assumption would be unjustified especially in the context
    of remote voting, where voting client programs (typically
    HTML/JS applications) run on voters' devices.  One cannot
    reasonably assume that such devices are properly maintained,
    patched and free of malware. Moreover, as often the code of
    the voting client application is served by the voting server,
    such trust assumption would have to be extended to such
    servers as well.

    The problem of providing adequate and usable solutions for
    cast-as-intended verifiability has recently attracted
    significant attention. In particular, various solutions based
    on the idea of \emph{return codes} have been proposed
    \cite{Allepuz11}, \cite{Gjosteen11}, \cite{Puigalli12},
    \cite{Gjosteen13}, \cite{GalindoGP15},
    \cite{Haenni16,HaenniKLD17}, where different solutions
    provide different balance of security and usability features.
    Notably, solutions based on return codes
    \cite{Gjosteen11,Gjosteen13,StenerudB12} were used in Norway
    in legally binding municipal and county council elections in
    2011 and 2013, \cite{GalindoGP15} was used in 2015 in binding
    elections in the Swiss canton of Neuch\^atel,
    while \cite{Haenni16}, as stated in the paper, is planned
    to be used as a part of the electronic voting system for the
    State of Geneva (Switzerland) \cite{HaenniKLD17}.

    The above mentioned solutions share the following underlying
    idea.  In the registration phase, each voter obtains over a
    trusted channel a ballot sheet, where pre-generated return
    codes (typically short random alpha-numeric sequences) are
    printed next to each voting choice. Then, in the voting
    phase, after the voter has selected her voting choices and
    the voting client application has submitted an encrypted vote
    to the remote voting server, the voting authorities
    compute/retrieve (in some way dependent on the specific
    solution) return codes which are meant to correspond to the
    choices made by the voter. These codes are sent back to the
    voter (possibly using an independent channel) who compares
    them with the codes printed on her ballot sheet next to
    the selected choices. The idea here is that when this match
    succeeds, the voter can be sure that the submitted
    (encrypted) vote indeed contains her intended choices (as
    otherwise the voter would not have obtained the matching
    codes). The voter may then finalize the ballot casting
    process (by, for instance, submitting some kind of
    finalization code) or, if the match does not succeed, she
    may undertake different steps (for instance, vote from
    another device or use a completely different voting method).

    \mpar{Our Contribution}

    In this paper we propose a new cast-as-intended mechanism
    based on return codes. Our solution provides an
    attractive balance of security and functional properties:
    \begin{enumerate}
    \item
        It is based on well-known cryptographic building blocks
        and relies on standard cryptographic assumptions, which
        allows for relatively simple security proofs. In fact,
        our proofs are modular in that they do not depend on the
        details of the underlying voting protocol to which our
        return code scheme is added. In particular, our formal
        security results hold for systems obtained
        by adding our return codes to the mix-net-based variants
        of the well-known Helios and Belenios protocols.
    \item
        Our scheme is designed with distributed trust in mind:
        the computations carried out to retrieve/compute return
        codes are distributed in their nature, such that a
        threshold of trustees must be corrupted in order to carry
        out a successful attack and fool the voter.
    \item
        Our solution imposes only very little additional
        computational burden on the voting platform, which is
        especially important if voters use devices of restricted
        computational power such as mobile phones. The
        computational cost on the server side is very reasonable
        and scales well with the increasing ballot size (it
        is, up to ballots of fairly big size, essentially
        constant).
    \end{enumerate}
    

    Our scheme is meant to provide \emph{cast as intended}
    verifiability even if the voting platform is controlled by
    the adversary under the following
    assumptions. First, we assume that not more than $t-1$
    tellers are corrupted (i.e.\ controlled by the adversary),
    where $t$ is the threshold of the used threshold-decryption
    scheme. Second, we assume that the printing facility and the
    ballot delivery channel are not corrupted. Under these
    assumptions, if the voter during the voting process obtains
    the expected return codes (that is the codes printed on her
    ballot sheet next to her intended choices), then the cast
    ballot is guaranteed to contain the intended voter's choice.

    We note that the second assumption is shared with other
    return code solutions. It is a strong assumption and requires
    special measures in order to be justified in specific
    deployments.
    The same assumption (in addition to the standard assumptions
    that the voter platform is honest and that at most $t-1$
    tellers are corrupted) is necessary for 
    \emph{voters' privacy}.
    Finally, note that our scheme (similarly to most of the return
    code solutions; see below for more discussion) is not meant
    to provide \emph{receipt freeness}.
    
    \medskip

    On the technical level, our scheme is inspired by the PGD
    system
    \cite{RyanTeague-WSP-2009,HeatherRyanTeague-ESORICS-2010}
    which however does not implement the idea of returns codes,
    but instead the one of \emph{voting codes} (where a voter
    \emph{submits codes} corresponding to her choice).  Sharing
    some similarities with this construction, our system differs
    substantially from PGD in many aspects. 

    As an additional contribution of this paper, we demonstrate
    an attack on a return code scheme recently proposed in
    \cite{Haenni16,HaenniKLD17} which was planned to be used in
    the context of the Geneva Internet voting
    project (see below for more details).


\mpar{Related Work}

    As already mentioned, our scheme is inspired by the PGD system
    \cite{RyanTeague-WSP-2009,HeatherRyanTeague-ESORICS-2010}
    and, on the technical level, uses some similar ideas: it uses
    distributed PETs (plaintext equivalence tests) to match the
    submitted ballots against a pre-published encrypted code
    table. Our scheme, however, differs from PGD in some
    significant ways.  
    Our scheme scales better with increasing
    ballot complexity (PGP performs one PET for every entry in
    the voter's code table; we perform only one PET per voter
    even for relatively complex ballots).
    On the technical level we avoid the use of
    encrypted permutations (onions). 
    Finally, PGD uses the idea of voting codes, where a voter
    submits codes corresponding to the chosen candidates
    (although the authors also suggest the possibility of using
    return codes). We note here that the use of voting codes (as
    in PGD) results in stronger ballot secrecy (the voting client
    does not get to learn how the voter's choice and hence it
    does not have to be trusted for ballot secrecy). As a
    trade-off, using voting codes tends to be less convinient for
    the voters.

    In a series of results including 
    \cite{Allepuz11,Puigalli12,Gjosteen11,Gjosteen13},
    related to the Norwegian Internet voting projects
    (\emph{eValg2011} and \emph{eValg2013}) \cite{StenerudB12},
    the underlying, shared idea is as follows. The code for a voting
    option $v$ (which is cast in an encrypted form $\enc{pk}{v}$) is deterministically derived from $v$ using a
    per-voter secret $s$ (it typically is $v^s$). This derivation
    process is carried out by two servers (playing fixed,
    specific roles) in such a way that if only one of them is
    corrupted, the security goal of the return codes is not
    subverted. 
    In order to make this idea work for more complex ballots,
    \cite{Gjosteen11,Gjosteen13} uses a technique of combining
    codes, which however requires some non-standard cryptographic
    assumption (hardness of the SGSP problem, where SGSP stands
    for \emph{Subgroup Generated by Small Primes}).  These
    schemes (as opposed to ours) do not allow for more
    fine-grained distribution of trust: there are exactly two
    parties with specific roles, one of which must be honest.

    The above idea was further transformed in a scheme proposed
    for the voting system in the canton of Neuch{\^{a}}tel in
    Switzerland \cite{GalindoGP15}, with the main technical
    difference that in this system a voter holds one part of the
    secret used for code generation (which causes some usability
    issues which were addressed by introducing of a so-called
    usability layer, which unfortunately weakens security
    guarantees). Security of this construction relies on the same
    non-standard security assumption as
    \cite{Gjosteen11,Gjosteen13} do and, similarly, there is no
    built-in fine grained mechanism for distributed trust.
    Compared to our system, this system requires much more
    complex computations on the voting platform, but less
    computations for the election authorities (although in both
    cases the ballot processing time on the server side is 
    essentially constant independently of the
    number of voting options).

    Recently, an interesting solution has been proposed in the
    context of the Geneva Internet voting project
    \cite{Haenni16,HaenniKLD17}. This solution is based on
    oblivious transfer, where, intuitively, the security of the
    mechanism is provided by the fact that the authorities (even
    although they may know all the codes) do not know which codes
    are actually transfered to the voter. This provides some
    level of protection against vote buying schemes which
    otherwise could be very easily mounted by a dishonest
    authority (if a voter was willing to disclose her ballot
    sheet). To our knowledge, this is the only return-codes
    scheme with this property.

    As a downside, in this protocol codes cannot be transfered
    using an independent channel (they must be transfered via the
    voter's platform), which rules out the use of this protocol
    in elections where re-voting is allowed.
    Furthermore, this protocol, again, uses the same non-standard
    cryptographic assumption as \cite{Gjosteen11,Gjosteen13}.

    Finally, as already mentioned, we have discovered a serious
    flaw in this construction, described in detail in
    Appendix~\ref{sect:attack}. Our attack violates the
    cast-as-intended property of the scheme (the voter cannot be
    sure that the cast ballot represents her intended choice even
    if she receives the expected return codes) and can be mounted
    by an attacker who only controls the voting platrofm. In
    short, we show that such an attacker (which is exactly the
    kind of attacker the system is meant to defend against)
    \emph{can cast invalid ballots} and still provide the voters
    with valid return codes.  These invalid ballots are accepted
    by the voting server, tallied, and only discovered and
    rejected after tallying, when the link between the ballot and
    the voter has been hidden.  Note that even if augmented the
    protocol with a mechanism enabling us to trace the
    mallformed decrypted ballots back to the voters, it would
    only point to dishonest voters' devices which cannot be held
    accuntable. 
    
    While there is a natural countermeasure for this attack
    (adding appropriate zero-knowledge proofs of well-formedness
    of the ballot), it comes with significant degradation of
    performance: it works, roughly, in quadratic time with
    respect to the number of voting options, which renders this
    solution impractical for bigger ballots.%
    \footnote{We contacted the authors who confirmed the flaw
        and are working on a countermeasure for the
        attack. According to the authors (personal communication,
        May 15, 2017), they were able to find a solution which
        avoids quadratic time computations \emph{during the
        ballot casting}, assuming instead quadratic-time
        pre-computations carried out by servers in the off-line phase.}
    
\mpar{Structure of the paper}

    After introducing some preliminary definitions in
    (Section~\ref{sec:preliminaries}) and providing an overview
    of the election process (Section~\ref{sec:overview}), we
    describe in Section~\ref{sec:simple} a simple variant our
    scheme, applicable only for ballots with one binary choice.
    The general variant is described in
    Section~\ref{sec:general}, after which the security analysis
    is presented in Section~\ref{sec:security}. In appendices we
    provide some further details and demonstrate the mentioned
    attack on \cite{Haenni16}.
     

\section{Preliminaries \label{sec:preliminaries}}

    Our return code scheme uses the well-known ElGamal
    cryptosystem over a cyclic group $G$ of quadratic residues
    modulo a safe prime $p = 2q + 1$. This cryptosystem is
    multiplicatively homomorphic (that is $\enc{pk}{m} \cdot
    \enc{pk}{m'}$ results in an encryption $\enc{pk}{m \cdot m'}$
    if $m$ and $m'$ are elements of the underlying group).  
    A distributed key generation protocol for the ElGamal
    cryptosystem (where $n$ tellers jointly generate a secret key
    and the corresponding public key, and pre-determined
    threshold $t<n$ out of $n$ tellers is necessary for
    decryption) is proposed, for instance, in
    \cite{GennaroEtAl-JKC-2007}.  

    A plaintext-equivalence test
    \cite{JakobssonJuels-ASIACRYPT-2000} is a zero-knowledge
    protocol that allows the (threshold of) tellers to verifiably
    check if two ciphertexts $c$ and $c'$ contain the same
    plaintext, i.e.\ to check if $\dec{sk}{c}= \dec{sk}{c'}$, but
    nothing more about the plaintexts of $c$ and $c'$.


    Our return codes solution can be added to any voting system
    with encrypted ballot of a form which is compatible with our
    scheme in the following sense: (1) ElGamal cryptosystem with
    threshold decryption, as introduced above, is used to encrypt
    voters' choices and (2) ballots contain zero-knowledge proofs
    of knowledge of the encrypted choices (which is a very common
    case); additionally, for the general case, we require that
    (3) voters' choices are encoded in a specific way (see
    Section \ref{sec:general}) before encryption.  We do not fix
    details of the authentication mechanism nor those of the
    tallying process. In fact, our security proofs work
    independently of these details.  Examples of voting systems
    compatible with our scheme are Helios
    \cite{Adida-Security-2008} and Belenios \cite{CortierGGI14}
    with mix-net-based tallying and, for the simple variant, also
    with homomorphic tallying.


    \section{Overview of the Election Process \label{sec:overview}}

    In this section we present an overview of the voting process.
    Because our scheme (like other return codes solutions) is
    aimed at providing cast-as-intended verifiability even when
    the voting platform is potentially corrupted, we make the
    distinction between \emph{voters} and their \emph{voting
    platform}, that is devices, including the software
    potentially served by the voting server, voters use to cast
    ballots.

    The election process is run by the set of
    \textbf{authorities} including:
    \begin{itemize}
    \item 
        \emph{Tellers} who jointly generate the public election
        key $\pk_e$ key and share the corresponding decryption
        key in a threshold manner. They also, similarly, jointly
        generate the \emph{public code key} $\pk_c$ which will be
        used to encrypt codes in code tables and an
        auxiliary public key $\pk_a$ for which the corresponding
        secret key is known to every teller (here we do not need
        threshold decryption and use any CCA2-secure
        cryptosystem). The tellers take part in code table
        generation and generation of additional codes for voters
        (authentication, finalisation and confirmation codes).
        They may also carry out additional steps (such as ballots
        shuffling), as specified by the underlying protocol.
    \item
        \emph{Secure bulletin boards} which, traditionally for
        e-voting systems, are used by voting authorities to
        publish results of various steps of the election
        procedure, including the final election result. Secure
        bulletin boards provide append-only storage, where
        records can be published (appended) but never changed or
        removed.
    \item
        \emph{Voting server} which is responsible for voters'
        authentication and ballot recording (where a ballot is
        published on a designated secure bulletin board).
    \item
        \emph{Printing facility}, including the ballot sheets
        delivery, used to print ballot sheets in a trusted way and to
        deliver ballot sheets to eligible voters. The printing
        facility, in the setup phase generates its private/public
        encryption key pair and publishes the public key $\pk_p$.
    \end{itemize}

    Our return code schemes supports the following, general
    \textbf{ballot structure}: a ballot may contain a number of
    voting options (candidates), where a voter can independently
    select each of these options (or, put differently, provide
    `yes'/`no' choice independently for each voting option).
    Further restrictions can be imposed (such as for example,
    that exactly $k$ or at most $k$ options are selected) and
    checked after the ballots are decrypted. Note that with this
    ballot structure we can encode different types of ballots,
    such as for instance, ballots where each candidate can get
    more than one vote. 

    The election process consists of the following \textbf{voting
    phases}:
   
    In the \emph{setup phase} the tellers and the printing
    facility generate keys and codes, as described above. 
    In the \emph{registration phase} every eligible voter obtains
    (via a trusted channel) a ballot sheet. The ballot sheet
    contains an \emph{authentication code} (used as a
    authentication measure; we abstract here from the details of
    the authentication mechanism and simply assume that a
    mechanism with sufficient security level is used), a
    \emph{finalization code}, a \emph{confirmation code}, and a
    list of voting options (candidates) with printed next to each
    of them two return codes: one for the `no' choice and one for
    the `yes' choice.

    In the \emph{voting phase}, the voter, using her voting
    platform and the authentication code, authenticates to the
    voting server and selects her choices. The voting platform
    creates a ballot with the selected choices and submits it to the
    voting server. The ballot is then processed by the voting
    authorities who send back to the voter (via the voting
    platform or via some other, independent channel) sequence of
    return codes that correspond to the cast (encrypted) choices.
    The voter compares the obtained codes with the ones printed
    on her ballot sheet to make sure that they indeed correspond
    to her intended choices. If this is the case, the voter
    provides the voting platform with the finalization code which
    is forwarded to the voting server.  Given this finalization
    code, the voting server sends the confirmation code to the
    voter and completes the ballot casting process by adding the
    ballot to the ballot box. If something does not work as
    expected (the voter does not get the expected return codes or
    does not obtain the confirmation code after providing her
    finalisation code), the voter can undertake special steps, as
    prescribed by the election procedure (use, for instance,
    another device or the conventional voting method).

    Finally, in the \emph{tallying phase}, the ballots published
    on the ballot box are tallied and the result is computed.


\section{The Variant with one Binary Choice \label{sec:simple}}

    In this section, we present a  simple variant of
    our scheme, where the ballot contains only one binary choice
    (two candidate races or `yes'/`no' elections). This variant,
    while avoiding the technical details of the general variant,
    demonstrates the main ideas of the scheme. 

    \paragraph{Code table and ballot sheet.}

    As shortly mentioned before, in the setup phase, the voting
    authorities generate for every voter an encrypted \emph{code
    table}. We will now only describe the expected result of the
    code generation procedure, without going into the detail. Such
    details will be given in Section~\ref{sect:code-generation},
    where the general case is covered (which subsumes the
    simple case discussed in this section). We only mention here 
    that code tables are generated in fully verifiable way.

    The code generation procedure generates, for every voter, two
    random codes $c_0$ and $c_1$, corresponding to the `no' and
    `yes' choice, and a random bit $b$, called a \emph{flip bit}.
    It also generates for every voter a random \emph{finalization
    code} and a \emph{confirmation code}. Additionally, we assume
    that some kind of \emph{authentication codes} for voters may
    be generated by this procedure as well, but we abstract away
    from the details of the authentication mechanism, as the
    presented construction does not depend on them. 

    The \emph{ballot sheet} (delivered to the voter over a
    trusted channel) contains the authentication, finalization,
    and confirmation codes, the return codes $c_0$ and $c_1$
    printed in clear next to, respectively, the `no' and the
    `yes' voting choice, and the flip bit $b$.  For usability
    reasons, the flip bit can be integrated into the
    authentication code, so that the voter does not have to enter
    it separately.

    The \emph{code table} associated with the voter, published on
    a bulletin board, is of the form
    $$
        \cfin, \, \econf, \, (e_0, d_0), (e_1, d_1)
    $$ 
    where $\cfin$ is a commitment to the finalization code,
    $\econf$ is encryption of the confirmation code under $\pk_c$
    and
    \begin{align*}
        &e_0 = \eenc{b}, 
        &&d_0 = \cenc{c_b},
        &e_1 = \eenc{1-b},
        &&d_0 = \cenc{c_{1-b}}.
    \end{align*}
    Note that the this record contains the pair of ciphertexts
    corresponding to the `no' choice (encrypted $0$ and encrypted
    code $c_0$) and the pair of ciphertexts corresponding to the
    `yes' choice (encrypted $1$ and encrypted code $c_1$).  The
    order in which these two pairs are placed depends on the flip
    bit (if the flip bit is $1$ the order is
    flipped).\footnote{Note that the plaintext are first mapped
    into $G$ before being encrypted; for an appropriate choice of
    the mapping, we obtain a system which coincides with the general
    variant with $k=1$ and, furthermore, allows for homomorphic tallying.}

    \paragraph{Ballot casting.}
    The voter provides her voting application with her
    authentication code, the flip bit $b$, and her voting
    choice $v \in \{0,1\}$. The voting application produces a
    ballot containing
    $$ 
        w = \eenc{v}, \ \ \aenc{\tilde b}, \ \ \pi
    $$
    where $\tilde b = v \oplus b$ and $\pi$ is a zero-knowledge
    proof of knowledge of the plaintext in the ciphertext $w$
    ($\tilde b$ is encrypted in order to hide it from an
    external observer; the tellers will decrypt this value in the
    next step).

    The voting authorities check the zero-knowledge proof $\pi$,
    decrypt $\tilde b$, select $e_{\tilde b}$ from the voter's
    table and perform the PET of this ciphertext with the
    ciphertext $w$ submitted by the voter's platform. It
    is expected that this PET succeeds (which is the case if the
    voting platform follows the protocol and the ballot sheet and
    the code table are correctly generated). If this is the case,
    the corresponding encrypted code $d_{\tilde b}$ is decrypted
    (which should result in $c_v$) and delivered to the voter.
    The voter makes sure that, indeed, the return code is $c_v$,
    i.e.\ it corresponds to the voting choice $v$, before she
    provides her finalization code (in order to finalize the
    ballot casting process). The voting authorities check that
    the provided finalization code is a valid opening for the
    commitment $\cfin$. If this is the case, they finalise the
    ballot casting process: they jointly decrypt the confirmation
    code, send it to the voter, and add the voter's ballot to
    the ballot box.
    
\paragraph{Tallying.}
    Finally, after the voting phase is over, ballots collected in
    the ballot box are tallied.
    We abstract here from the details of the tallying procedure.
    Importantly, our security results work regardless of the
    details of this procedure.

\medskip\noindent
    The intuition behind security of this scheme is as follows.
    Because, of the correctness of the code table and PET
    operations (which is ensured by zero-knowledge proofs), if
    the PET succeeds, then the decrypted code must be the return
    code corresponding to the \emph{actual} plaintext in the
    encrypted ballot. To fool the voter, an adversary would have
    to send him the code contained in the second ciphertext which
    has not been decrypted. But the best the adversary can
    do---not being able to break the used encryption scheme---is blindly
    guess this code, which gives him very small probability of
    success.
    
    \begin{remark}
        For this simple variant, we do not really need to include
        the flip bit in the ballot sheet: the ciphertext $w$
        could be matched, using the PET protocol, against both 
        $e_0$ and $e_1$, one of which should succeed, which would
        determine $\tilde b$. Including the flip bits in the
        ballot sheets is however crucial for efficiency
        of the general variant. 
    \end{remark}

    \noindent
    We can note that the additional computational cost of this
    scheme added to the voting platform is only one encryption.
    The computational cost incurred by this scheme on the server
    side (per one voter) is one additional decryption to decrypt
    $\tilde b$, one verifiable PET, and one distributed
    decryption to decrypt the return code.


    As we will see in a moment, the general variant of our scheme
    (with $k$ independent choices) can be seen as a
    combination of $k$ simple cases as described here with some
    optimisations. Interestingly, with these optimisations, the
    additional computational cost incurred by our scheme---if the
    size of the ballot does not grow too much---remains
    essentially the same.


\section{The General Variant}\label{sec:general}

    In this section we present the general variant of our code
    voting scheme, where ballots can contain some number $k$ of
    independent binary choices, one for each voting option. This
    variant is expressive enough to handle wide variety of complex
    ballots. Despite some technical details used for
    optimisation, this variant shares the same underlying idea,
    illustrated by the simple variant.

    We assume some encoding $\gamma$ of the voting
    options $1, \dots, k$ as elements of the group $G$ such that
    the voter's choice, which is now a subset of individual
    voting options, can be encoded as the multiplication of the
    encodings of these individual options. Of course, we assume
    that the individual voting options can be later efficiently
    retrieved from such an encoding. As an example of such
    encoding we can use the technique used for instance in
    \cite{GalindoGP15,Haenni16}, where the voting options are
    encodes as small prime numbers which belong to the group $G$. 
    
    Similarly, we assume a family of efficient encodings
    $\delta_i$ ($i \in \{1,\dots,k\}$) from the set of return
    codes to the group $G$, such that individual codes
    $c_1,\dots,c_k$ can be efficiently extracted from the product
    $\delta_1(c_1)\cdot \dots \cdot\delta_k(c_k)$. An example of
    such an encoding is given in
    Section~\ref{sect:encoding-delta}.

\subsection{Ballot Structure and Voting Procedure}

\paragraph{Code table and ballot sheets.}
    The code generation procedure is described in details in
    Section~\ref{sect:code-generation}. In addition to 
    finalisation and confirmation codes which are generated
    as previously, this procedure generates, for every voter and
    every voting option $i \in \{1, \dots, k\}$, two random codes
    $c^0_i$ and $c^1_i$ corresponding to, respectively, the `no'
    and `yes' choice. It then generates a random sequence of flip
    bits $\vec b = b_1,\dots,b_k$, where $b_i \in \{0,1\}$. 
    
    The ballot sheet sent to the voter contains now, besides the
    authentication, finalisation, and confirmation codes, return
    codes $(c^0_1, c^1_1), \dots, (c^0_k, c^1_k)$ printed in
    clear next to corresponding voting options and marked as,
    respectively the `no' and the 'yes' choice. It also contains
    the flip bits $\vec b$ (as before, this vector can be
    integrated in the authentication code).
    
    The published code table associated with the voter contains, 
    as before $\cfin$, $\econf$ and
    $$
        \bigl( u_i^0, u_i^1  \bigr)_{i=0}^k = 
        \bigl( t_i^{b_i}, \ t_i^{1-b_i} \bigr)_{i=0}^k
    $$
    where 
    $$
        t_i^0 = (\eenc 1,\, \cenc{\delta_i(c_i^0)}
        \quad\text{ and }\quad
        t_i^1 = (\eenc{\gamma(i)},\, \cenc{\delta_i(c_i^1)}.
    $$  
    Note that $t_i^0$ corresponds to the `no' choice (it contains
    an encryption of $1$ and the encoded code for `no') and
    $t_i^1$ corresponds to the `yes' choice (it contains an
    encryption of the encoded option $i$ and the encoded code for
    `yes').  Note also that $u_i^{b_i} = t_i^0$ and  $u_i^{1 -
    b_i} = t_i^1$.

\paragraph{Ballot casting.}
    The voter provides her voting application with her voting
    choice $v_1, \dots, v_k \in \{0,1\}$ and the bit sequence
    $\vec b$. The voting application computes $v = 
    \prod_{i \in V} \gamma(i)$, where we define $V$ as the set
    $\{j : 1\leq j \leq k,\; v_j = 1\}$,
    and produces a ballot containing
    $$ 
        w = \eenc{v},\ \  \aenc{\tilde{\vec b}}, \ \ \pi
    $$
    where $\pi$ is, as before, a zero-knowledge proof of
    knowledge of the plaintext of $w$ and
    $\tilde{\vec b} = \tilde b_1, \dots, \tilde b_k$ with
    $\tilde b_i = b_i \oplus v_i$.

    The voting authorities decrypt $\tilde{\vec b}$ and select 
    the values $w_i = u_i^{\tilde b_i}$, for $i \in \{1,\dots,k\}$.  
    Note that if the voter has \emph{not} chosen the $i$-th
    election option, then $w_i = u_i^{b_i} = t_i^0$, by the
    definition of $u$. Otherwise, $w_i = u_i^{1-b_i} = t_i^1$.

    The voting authorities multiply $w_1, \dots, w_k$
    (component-wise) obtaining the pair $(e^*, c^*)$, where
    $e^*$ should be (if the voter platform followed the
    protocol) encryption of $v = \prod_{i\in V} \gamma(i)$. The
    voting authorities perform the PET of $e^*$ with the
    encrypted choice $w$ from the ballot. If this PET fails,
    the casting procedure is canceled. Otherwise, the decryption
    tellers jointly decrypt $c^*$.  Observe that, by the
    properties of the published code table, this decrypted value
    is the product of $\delta_j(c_j^{v_j})$, i.e. it is the
    product of the codes corresponding to the choices made by the
    voter. This value is decomposed 
    into individual codes $c_1^{v_1}, \dots c_k^{v_k}$ and
    sent to the voter (via the voting platform or an independent
    channel). As before, the voter makes sure that the received
    codes correspond to her choices before providing the
    finalisation code.

    \medskip\noindent
    Note that the ballot processing on the server side only
    requires one verifiable PET, one decryption and one threshold
    decryption, independently of the number $k$ of the voting
    options, plus some number of multiplications and divisions
    (which depends on $k$), as long as $k$ codes can be
    efficiently represented as one element of the group $G$ which
    is in detail discussed in Section~\ref{sect:encoding-delta}.

\subsection{Code Table Generation\label{sect:code-generation}}

    The code table generation presented below is fully
    verifiable.  Note that we could also consider a version
    without zero-knowledge proofs, but with partial checking
    instead, where a bigger number of records is produced and the
    some of them (randomly selected) are open for audit.

    We will assume that the code generation procedure is carried
    out by the tellers, but it can by carried out by any set of
    independent parties, as it does not require possession of any
    secret keys. We will present here a version, where, for the
    same voting option, distinct voters obtain distinct codes,
    although different variants are also possible (and may be
    useful if the number of voters is very big).

    The set of codes is $\codes = \{1,\dots,m\}$ with $m > 2n$,
    where $n$ is the number of voters (reasonable values for $m$,
    that is values corresponding to desired security levels, 
    can be determined using Theorem \ref{th:cai}).

    For simplicity of presentation, in the following, we will
    leave out handling of the authentication, finalization and
    confirmation codes.  The procedure consists of the following
    steps.

    \begin{enumerate}
    \item            
        For every voting option $j$, the tellers
        deterministically compute
        $$
            \cenc{\delta_j(1)}, \penc{1} ,
            \dots,
            \cenc{\delta_j(m)}, \penc{m}.
        $$
        where all the ciphertext are obtained using the
        pre-agreed randomness $1$.
    \item
        The tellers shuffle the above sequence of ciphertexts
        using a verifiable mix net obtaining a sequence of the form
        $$
        \cenc{\delta_j(c_1)}, \penc{c_1} ,
        \dots,
        \cenc{\delta_j(c_m)}, \penc{c_m}, 
        $$
        where $c_i = \pi(i)$ for some permutation $\pi$ and the
        ciphertext are re-randomized.  Note that for this we need
        to use a version of verifiable mixing which applies the
        same permutation (but independent re-randomization
        factors) to pairs of ciphertexts. Such generalizations of
        know verifiable shuffling algorithms are
        possible.\footnote{In particular, it is straightforward
        to generalize the shuffle protocol of
        \cite{BayerGroth-EUROCRYPT-2012} to provide such
        functionality.}

    \item
        The tellers take the consecutive encrypted codes produced
        in the previous step and organize them into the records
        of the following form, one for each voter $i$:
        \begin{align*}
            \bigl\{ \ 
                &\penc 0, \penc{c'_j}, 
                \penc 1, \penc{c''_j}, \\
                &\eenc{1}, \cenc{\delta_j(c'_j)}, 
                \eenc{\gamma(j)}, \cenc{\delta_j(c''_j)} 
            \ \bigr\}_{j\in\{1,\dots,k\}}
        \end{align*}
        where  the ciphertext with (encoded) choices are
        generated deterministically with the randomness
        $1$.
        
    \item
        The tellers perform, one after another, series of
        micro-mixes for every such a record: Each teller, for the
        input record $R = (a_1, b_1, a_2, b_2, a'_1, b'_1, a'_2,
        b'_2)$ (which is the output of the previous teller or,
        for the first teller, the record produced in the previous
        step) picks a random bit.  If this bit is $0$, then it
        only re-encrypts all the elements $R$. If the flip bit is
        $1$, then, in addition, it accordingly flips the elements
        of the record and outputs a re-encryption of $R' = (a_2,
        b_2, a_1, b_1, a'_2, b'_2, a'_1, b'_1)$. The teller
        produces a zero-knowledge proof of correctness of this
        operation (such step can be implemented as a verifiable
        mixing operation; it can be also realized using
        disjunctive Chaum-Pedersen zero-knowledge proofs 
        of the fact that the resulting
        record is either a re-encryption of $R$ or $R'$).

    \item
        The parts of the records encrypted with $\pk_c$ and
        $\pk_e$ are published in voters' code tables. The parts
        encrypted with $\pk_p$ are given to the printing facility
        which decrypts the records. The decrypted content
        contains the return codes and (implicitly, via the order
        of plaintexts) the flip bit sequence $\vec b$.
    \end{enumerate}

    Because, in the above procedure, all the operation are fully
    deterministic or with appropriate zero-knowledge proofs, we
    obtain the following result.

    \begin{theorem}\label{th:codes}
        The above procedure is corrects, i.e.\ it produces
        correctly linked ballot sheets and encrypted code tables
        with overwhelming probability. Moreover, unless the
        threshold of trustees are dishonest, only the printing
        facility learns how codes are distributed amongst voters.
    \end{theorem}


    \section{Security Analysis \label{sec:security}}

    As noted in the introduction, coercion resistance and
    receipt-freeness are not the goals of our scheme. In fact,
    the use of return codes, as in many similar solutions,
    specifically makes the scheme prone to vote selling \emph{if
        dishonest authorities are involved in the malicious
    behaviour}.

    The results presented in this section are stated for the case
    where re-voting is not allowed. For the case with re-voting
    (casting multiple ballots, of which, say, the last is
    counted), we expect that the privacy result holds (we leave
    however the proof for future work), while only a weaker form of
    cast-as-intended verifiability than the one presented in
    Section \ref{sect:cai} can be guaranteed: namely, we have to
    assume that an independent channel is used to send return
    codes to voters and that both the tellers (who see the sent
    return codes) and this channel are honest. 

\subsection{Ballot Secrecy}

    Ballot secrecy means, informally, that it is impossible (for
    an adversary) to obtain more information about the choices of
    individual \emph{honest} voters (that is voters following the
    protocol), than can be inferred from the explicit election
    result.  Our code voting scheme provides voters privacy under
    the following assumptions:
    \begin{enumerate}[P1.]
    \item
        The voting platform is not corrupted. 
    \item
        At most $t-1$ tellers are corrupted, where $t$ is the
        threshold for decryption.
    \item
        The printing facility and the ballot sheet delivery
        channel are not corrupted.
    \end{enumerate}
    The first two assumptions are standard and for voters'
    privacy and shared by many e-voting protocols (using and not
    using return codes). The third assumption is also shared by
    any code voting scheme (where codes need to be printed and
    delivered to the voter).  Therefore, in this sense, these are
    the minimal assumptions for electronic voting with return
    codes.

    Note also, that the informal definition of privacy given
    above only protect honest voters who, in particular, do not
    reveal their ballot sheet to another parties, excluding
    voters who want to sell their ballots.

    We formalize the above notion of privacy using the following
    game between the adversary and the system $P$ representing
    the honest components of the e-voting system, where the
    adversary gets to control all but two (honest) voters. For
    simplicity of presentation, we consider here the simple case
    where voters have only one yes/no choice.  We will consider
    two variants of $P$: variant $P_0$, where the first of the
    honest voters votes for the `no' option and the second honest
    voters chooses the `yes' option, and variant $P_1$, where the
    choices of the honest voters are swapped.  With these
    definitions, we express the notion of privacy by requiring
    that there is no polynomially bounded adversary $A$ which can
    detect if he is interacting with  $P_0$ or $P_1$ (up to some
    negligible probability), that is:
    \begin{equation}\label{privacy}
        \prob[P_0 \comp A \mapsto 1] \equivnegl
        \prob[P_1 \comp A \mapsto 1]
    \end{equation}
    where $P_i \comp A \mapsto 1$ denotes the event that in the
    run of the system composed of $P_i$ and $A$, the adversary
    outputs $1$ (accepts the run).  We assume that the adversary
    can interact with the system in the following way: it
    interacts with the honest tellers playing the role of the
    dishonest tellers (in all the protocol stages). It also casts
    ballots of the dishonest voters and, at some chosen time,
    triggers the honest voters to cast their ballots. 

    We will now establish our privacy result in a modular way,
    independently of many details of the underlying voting system
    to which our return code scheme is added. We only assume that
    the system has the structure which allows for the described
    above game and which uses ballot encoding `compatible' with
    our construction, as described in
    Section~\ref{sec:preliminaries}.

    \begin{theorem}\label{th:privacy}
        Let $U$ be the underlying voting protocol and let $P$
        denote the protocol obtained from $U$ by adding our
        return code scheme. If $U$ provides ballot secrecy, as
        expressed by \eqref{privacy}, then $P$ provides secrecy
        as well.
    \end{theorem}

    In the proof we show that all the additional elements of $P$
    (related to codes) can be simulated by a simulator which has
    only black-box access to $U$. See Appendix \ref{sec:thproof}
    for more details.

\subsection{Cast-as-intended Verifiability\label{sect:cai}}

    Cast-as-intended verifiability means that an honest voter
    can, with high probability, make sure that the ballot cast on
    her behalf by her voting platform and recorded in the ballot
    box by the voting server contains her intended choice. 
    Our scheme provides cast-as-intended verifiability under one
    of the following cases:
    (1)
        The voter client is honest.
    (2)
        The following trust assumptions are satisfied:
        \begin{enumerate}[V1.]
        \item
            At most $t-1$ tellers are corrupted.
        \item
            The printing facility and the ballot sheet delivery
            channel are not corrupted.
        \end{enumerate}
    The first case is trivial (note that the very purpose of code
    voting is to provide cast-as-intended verifiability in the
    case the voter client is \emph{not} honest). We only need to
    assume that the voting client has means to check that the
    cast ballot has been in fact added to the ballot box and that
    there is a mechanism preventing any party from removing or
    modifying this ballot (which is the case if we assume that
    the ballot box is an instance of a secure bulletin board).

    In the following we analyse the second case. Our result is as
    follows.

    \begin{theorem}\label{th:cai}
        Under the assumption V1 and V2, for any given honest
        voter (possibly using a dishonest voting platform) and
        for any of the $k$ voting options, the probability that the
        voter obtains the expected code (that is the code printed
        next to voter's choices), while the recorded ballot
        contains different choices for this voting option is not
        bigger than $\frac{1}{m - n - n'}$ (plus a negligible
        value), where $m$ is the number of generated codes,
        $n$ is the total number of voters, and $n' < n$ is the
        number of corrupted voters.
    \end{theorem}

    The proof of this result, similarly to the privacy result,
    does not depend on the details of the authentication
    mechanism nor on the details of the tallying phase.
    
    The intuition behind this statement is that everything that
    the adversary can learn about code distribution is,
    essentially (up to cases of a negligible probability), what
    is explicitly given to him, that is (a) $n$ codes that have
    been decrypted by the tellers (b) $n'$ remaining codes of
    dishonest voters (because the adversary gets to see all the
    codes of these voters).  So, if the adversary wants to come
    up with the code corresponding to the opposite choice (for
    the considered voting option) of the honest voter in order to
    fool her, the best he can do is pick one of the remaining
    $m - n - n'$ codes at random. 
    
    In order to prove this statement we, similarly to the privacy
    proof, use the ideal (honest) code generation procedure and
    replace PETs by the appropriate ideal functionality. In this
    setting the following is true.

    Because the code table is correct (correctly corresponds to
    the printed ballot sheets) and the results of PETs is correct
    too (as we are using the ideal functionality), it follows
    that the decrypted codes correspond to the actual voting
    options in the encrypted ballots. We can then show that, if
    the adversary had a strategy of guessing an unencrypted code
    of an honest voter with better probability than given by the
    blind guess as described above (where the adversary picks one
    of the possible codes at random), this would break the
    IND-CPA property of the underlying encryption scheme.




\appendix

\section{Encoding for Codes\label{sect:encoding-delta}}
    
    In this section we describe some potential instantiations for
    the family $\delta_k$ ($i \in
    \{1,\dots,k\}$) of functions from the set of codes to the
    group $G$ such that individual codes $c_1,\dots,c_k$ can be
    efficiently extracted from a product $\delta_1(c_1)\cdot
    \dots \cdot\delta_k(c_k)$.

    A simple construction is to use the initial small prime
    numbers $p_0,p_1,\dots$ which belong to the group $G$ to
    represent consecutive bits of the binary representation of
    codes, as described below. Let us consider codes of the size
    of $l$ bits.

    For a code $c$ with the binary representation
    $b_0,\dots,b_{l-1}$, the function $\delta_k$ will use $l$
    primes, say, $p_{kl - l}, \dots p_{kl - 1}$ as follows:
    \begin{equation}\label{eq:primes}
        \delta_k(c) = \prod_{j\in\{0,\ldots,l-1\},\, b_j=1} 
            p_{kl - l + j}.
    \end{equation}
    That is, we take the product of those primes that corresponds
    to non-zero bits. Note that different $\delta_k$ use
    different primes. For the decoding (computing the inverse of
    $\delta_k$) one simply needs to factorize the resulting value
    into the used small primes, which can be done efficiently.
    Note that, since the considered primes are in $G$, the
    product \eqref{eq:primes} is in $G$ as well, provided that it
    is smaller than $p$. Moreover, because we want to multiply
    encrypted $\delta_k(c)$, we need to make sure that the
    product of such values does not grow beyond $p$ as well. This
    imposes a limit on how many codes can we
    represent in a single ciphertext.  For
    instance, for the standard 3072-bit ElGamal group, using this
    encoding we can have about 300 (exactly 296) independent
    bits, which allows us to handle 30 disjoint voting questions
    with 2-Base32-character codes, or 15 voting questions
    with 4-Base32-character codes. If a ballot sheet needs to contain
    more than this, additional
    ciphertexts would have to be used. 

    We can, however, relatively easy obtain denser encoding.
    If, for instance, instead of using one prime to represent one
    bit of a code, we use groups of 32 consecutive prime numbers to
    represent 5 bits of the code (where only one prime in the
    group is set), then the same group can fit around 1000 bits
    in one ciphertext. This gives 100 disjoint voting questions
    with 2-character codes or 50 questions with 4-character
    codes, a number corresponding to relatively complex ballots.

\section{Attack on \cite{Haenni16}\label{sect:attack}}    

    In order to understand the attack presented below, it may be
    useful for the reader to first consult the original paper
    \cite{Haenni16}.  It is worth noting that this attack
    scenario does not undermine the underlying ($k$ out of
    $n$)-OT scheme. It only utilizes the fact that a dishonest
    receiver in this scheme can obtain up to $k$ (but not more)
    values even if it does not follow the protocol.  We describe
    here an attack for the case with $n=k=2$.

    The \emph{intended run} of the protocol is as follows.
    For the voter's choice $\vec s = (s_1, s_2)$, the voting
    platform (VP)
    prepares an OT query 
    \[
        \vec a = (a_1,a_2), \quad\text{where}\quad 
        a_j = \Gamma(s_j) \cdot y^{r_j},
    \]
    for random $r_j$, where $y$ is the public election key. It
    also computes $b = g^{r_1 + r_2}$.
    Let $a$ denote the product of elements of $\vec a$ that is
    $a_1 \cdot a_2$. Note that $c = (b,a)$ is an ElGamal
    ciphertext (which, although not explicitly sent, will be
    considered to be the ciphertext cast by the voter) encrypting
    the plaintext $p = \Gamma(s_1) \cdot \Gamma(s_2)$ with
    randomness $r = r_1 + r_2$. The VP sends $\vec a$ and $b$
    along with a ZKP of knowledge of $r$ and $p$.

    From the OT response, the VP can now compute the codes for
    $s_1$ and $s_2$ which are shown to the voter who provides the
    confirmation code and the protocol goes on. Here are the
    details of how the codes are retrieved. The OT response
    contains:
    \begin{align*}
        & a_1^\alpha, \ a_2^\alpha, \ y^\alpha, \\
        & c_1 \oplus H(\Gamma(s_1)^\alpha), \ 
          c_2 \oplus H(\Gamma(s_2)^\alpha), \ \ldots
    \end{align*}
    for some random $\alpha$, where $c_1$ and $c_2$ are the codes
    corresponding to choices $s_1$ and $s_2$.  Knowing $r_1$ and
    $r_2$, the VP can compute $\Gamma(s_1)^\alpha$ and
    $\Gamma(s_2)^\alpha$ and, in turn, the codes $c_1, c_2$.

    \medskip

    The \emph{dishonest run} goes, for example, like this: For
    the voter's choice $\vec s = (s_1, s_2)$ as before, the VP
    prepares the OT query 
    \[
        \vec{\tilde a} = (a_1,\tilde a_2), \quad\text{where}\quad 
        \tilde a_2 = \Gamma(s_1)^7 \cdot \Gamma(s_2) \cdot y^{r_2}
    \]
    and sends $ \vec{\tilde a} $ along with $b$ and a ZKP of
    knowledge of $r$ and the plaintext $\tilde p$, which is now
    $\Gamma(s_1)^8 \cdot \Gamma(s_2)$. Jumping ahead, this
    plaintext will be rejected as invalid, but only after
    (mixing) and final decryption, when there is no visible link
    between the decrypted ballot and the voter.

    Nevertheless, from the OT response, the VT can easily compute
    the codes for $s_1$ and $s_2$ and make the protocol proceed
    as if the intended, valid ballot was cast. To see this,
    we can notice that, given the OT response, the VT can compute
    values $\Gamma(s_1)^\alpha$ and $(\Gamma(s_1)^{7} \cdot
    \Gamma(s_2))^\alpha$, from which it is easy to compute
    $\Gamma(s_2)^\alpha$ and the same codes $c_1$ and $c_2$ as in
    the honest run. These codes are delivered to the voter who
    then continues the procedure.

    \medskip\noindent
    A straightforward countermeasure for this attack would be
    adding appropriate zero-knowledge proofs of correctness of
    each $a_j$, which however adds a significant computational
    overhead (it works in time $O(k \cdot n)$).

\section{Proof of Theorem \ref{th:privacy}} \label{sec:thproof}

    We will now sketch the proof of this theorem, under the
    simplifying assumption that the code generation procedure is
    honest, i.e.\ carried out by one honest party (which is part
    of $P$). This assumption is justified by
    Theorem~\ref{th:codes}.

    First, we can observe that
    \begin{align*}
        \prob[ P_0 \comp A \mapsto 1] = \ 
        & \prob[ P_0 \comp A \mapsto 1 \wedge \vec b=(0,0) ] 
         + \prob[ P_0 \comp A \mapsto 1 \wedge \vec b=(0,1) ] + \\
        & \prob[ P_0 \comp A \mapsto 1 \wedge \vec b=(1,0) ] 
        + \prob[ P_0 \comp A \mapsto 1 \wedge \vec b=(1,1) ]
    \end{align*}
    where $\vec b$ represent the pair of flip-bits of the honest
    voters, and similarly for $P_1$. We can show that
    $
        \prob[P_0 \comp A \mapsto 1] = \prob[P_1 \comp A \mapsto 1]
    $ 
    by showing that
    \begin{align*}
        &\prob[P_0 \comp A \mapsto 1 \wedge \vec b=(0,0) ] 
        \equivnegl \prob[P_1 \comp A \mapsto 1 \wedge \vec b=(1,1) ]\\ 
        &\prob[P_0 \comp A \mapsto 1 \wedge \vec b=(1,0) ] 
        \equivnegl \prob[P_1 \comp A \mapsto 1 \wedge \vec b=(0,1) ] 
    \end{align*}
    and so on. We  focus on the first equation; the remaining
    cases are similar.

    In order to prove this equation, we construct a simulator $S$
    such that
    \begin{align*}
        & \prob[ P_0 \comp A \mapsto 1 \wedge \vec b=(0,0) ] 
         \equivnegl \prob[ U_0 \comp S \comp A \mapsto 1 
            \wedge \vec b=(0,0) ]
    \end{align*}
    and
    \begin{align*}
        & \prob[ P_1 \comp A \mapsto 1 \wedge \vec b=(1,1) ] 
         \equivnegl \prob[ U_1 \comp S \comp A \mapsto 1 
            \wedge \vec b=(1,1) ]
    \end{align*}
    and, furthermore, $S$ does not use $\vec b$ in any way
    (except that it picks this value at random). 
    This gives
    \begin{align*}
        & \prob[ P_0 \comp A \mapsto 1 \wedge \vec b=(0,0) ] 
          \equivnegl \prob[ U_0 \comp S \comp A \mapsto 1 
            \wedge \vec b=(0,0) ] \\
        & = \frac14\cdot \prob[ U_0 \comp S \comp A \mapsto 1 ] 
         \equivnegl \frac14\cdot \prob[ U_1 \comp S \comp A \mapsto 1 ] \\
        & = \prob[ U_1 \comp S \comp A \mapsto 1 \mid \vec b=(1,1)] 
          \equivnegl \prob[ P_1 \comp A \mapsto 1 \mid \vec b=(1,1) ].
    \end{align*}
    Which completes the proof.

    We construct $S$ in the following way. Recall that $S$ must
    simulate all the honest components of $P$ and during this
    simulation it can interact with $U$ (in the role of an
    adversary for $U$). 
    
    \paragraph{Setup.}
    $S$ forwards all the messages of the adversary related to the
    underlying protocol directly to $U$ and vice versa. It
    simulates the honest tellers in the procedures added in $P$
    (generation of additional keys). Finally, for code table and
    ballot sheets it simulates the process as prescribed by the
    protocol with the exception of the code table of the honest
    voters which are created as follows. 
    
    $S$ first triggers the honest voters (of $U$) to cast their
    ballots. Let $u_0$ and $u_1$ be the encrypted choices of
    these voters taken from these ballots. $S$ then randomly
    picks the flip bits for the honest voters but ignores their
    values (these values will not be used in the following).
    The code table entries of the honest voters, as computed by
    $S$, are respectively
    $$
        c, (u'_0, d_0), (u'_1, d'_0)
        \qquad\text{and}\qquad
        c', (u''_0, d_1), (u''_1, d'_1)
    $$
    where $c, c', d_0, d'_0, d_1, d_1'$ are generated like in the
    original code generation procedure (where, in particular the
    codes are picked at random), $u'_0, u''_0$ are
    (independent) re-encryptions of $u_0$ and, similarly $u'_1,
    u''_1$ are (independent) re-encryptions of $u_1$.

    This produces the code table with exactly the same distribution
    as the honest procedure (under the given condition
    restricting $\vec b$).

    \paragraph{Honest voters.}
    For the first honest voter $S$ uses the original ballot
    (containing $u_0$) as produced by the honest voter when it
    was triggered and adds to it the encryption of $\tilde b = 0$.
    Similarly for the second honest voter, but now $\tilde b$ is
    set to $1$.

    \paragraph{PETs.} 
    By the security properties of the PET protocol, $S$ can
    simulate its shares without knowing the secret key share, if
    only the result of PET known, which holds in our case: for
    the honest voters the result of PETs is simply `true'; for
    the dishonest voters the simulator can run the extraction
    algorithm to extract the used plaintexts from the
    zero-knowledge proofs included in the ballots, which allows
    the simulator to determine the result of the PETs, as it
    knows the plaintexts in the code tables of dishonest ballots
    (the simulator generated these plaintexts in the setup
    phase).

    \paragraph{Decryption of codes.} $S$ knows and uses its
    secret key share.

    \medskip\noindent
    Note that, indeed, in this simulation, $S$ does not make any use
    of the flip bits (except that it generates it).

\end{document}